\documentstyle[aps,multicol]{revtex}
\input epsf
\title
{
Singular Laplacian Growth
}
\author{Mark A. Peterson}
\address{
Physics Department, Mount Holyoke College, South Hadley MA 01075 USA
}
\date{October 4, 1997}
\begin{document}
\maketitle
\begin{abstract}
The general equations of motion for two dimensional singular Laplacian 
growth are derived using the conformal mapping method.  In the singular 
case, all singularities of the conformal map are on the unit circle, 
and the map is a degenerate Schwarz-Christoffel map.  The equations
of motion describe the motions of these singularities.  Despite the 
typical fractal-like outcomes of Laplacian growth processes, the 
equations of motion are shown to be {\em not} particularly sensitive 
to initial conditions.  It is argued that the sensitivity of this 
system derives from a novel cause, the non-uniqueness of 
solutions to the differential system.  By a mechanism of singularity 
creation, every solution can become more complex, even in the
absence of noise, without violating 
the growth law.  These processes are permitted, but are not
required, meaning the equation of motion does not determine the motion,
even in the small.
\end{abstract}
\pacs{PACS numbers: 68.70.+w, 02.30.Hq}

\section{Introduction}
\begin{multicols}{2}
Laplacian growth -- growth of a region  $D$ along the gradient of its external
Green's function -- is a model for a number of growth processes which
occur in nature, among them growth by electrodeposition \cite{sawada86}, 
diffusion-limited
aggregation \cite{witten81}, 
and viscous fingering at fluid-fluid interfaces \cite{nittmann85}.  These natural
processes exhibit very complicated morphologies, as does the mathematical
model.  
In spite of the large amount of work which has
been done, one still has the feeling that there is something
mysterious about Laplacian growth.  In particular, its extreme sensitivity to
perturbations makes it difficult to interpret experiments, real or numerical.

A special role is played in the two-dimensional problem by the conformal mapping method,
as developed in
\cite{paterson81},\cite{shraiman84},\cite{howison86},\cite{peterson89},\cite{blumenfeld95},
for example.  This method gives insight difficult to attain otherwise.   
It is especially simple
in radial geometry.  In outline, one
parametrizes the growing 2-dimensional region $D$, thought of as occupying
a bounded, simply connected region in the complex w-plane, by the conformal map
\begin{equation}
w=G(z)
\end{equation}
 which
takes the exterior of the unit disk, $|z|>1$, onto the exterior of $D$.  The
growth of $D$ is then represented by the time dependence of the conformal map $G$.
Since $G$ is holomorphic in $|z|>1$, and its dependence at infinity is also prescribed,
$G$ may, in turn, be parametrized by its singularities in the unit disk $|z|\leq 1$.
The growth becomes the dynamics of those singularities.
This method eliminates sources of inaccuracy 
which are unavoidable in other
methods, for example the statistical noise which accompanies DLA simulations,
or some of the roundoff and truncation errors of more straightforward integration methods.
That does not make the conformal mapping method necessarily more realistic, of course.
Indeed, the noise in other numerical methods may model actual physical noise, and
hence be desirable, if one's aim is to model particular examples of Laplacian-like growth.
With the conformal mapping method we aim rather to strip the problem down to its simplest
form, to see what remains and is common to all such processes.

In reference \cite{peterson89} this process of stripping down was taken one
level further, and an unexpected phenomenon came to light.  This case might be
called singular Laplacian growth, because it is the limiting case
in which all the singularities of $G$ are {\em on}
the unit circle $|z|=1$.  In this case $G$ degenerates to a Schwarz-Cristoffel map onto
the exterior of a
degenerate polygon $D$ of zero area (i.e., $D$ looks like a tree graph).  
In this limit the dynamics becomes 1-dimensional, and can be understood completely.
The surprise
is that the dynamics allows the singularities of $G$ to split and proliferate,
but it does not require this.  That is, while the dynamics is formally given by
differential equations, the solutions, for given initial data, are not unique.
The comment was made in Ref. \cite{peterson89} that this 
appears ``unlike any other physical model.''
In particular, it is not the same as being very sensitive to initial conditions,
as one might have assumed.  In fact, as we show below, 
singular Laplacian growth is not at
all sensitive in this way.  Its peculiarities have a different origin.

Reference \cite{peterson89} gave only 
the simplest example (in which all
maps could be written down explicitly), and not a general computable theory.
The present paper gives the general theory.

\section{Dynamics of Singularities}
Let $w=G(t,z)$ be a time-dependent conformal map of the exterior of the unit disk in the
$z$ plane onto the exterior of the domain $D$ in the $w$ plane.  G has the form
\begin{equation}
G(t,z)=z\sum_{k=0}^{\infty}c_k(t)z^{-k}.
\end{equation}
Suppose for the moment that $\partial D$, the boundary of $D$, is an analytic Jordan curve so
that there is no difficulty in defining 
\begin{equation}
g(t,\theta)=G(t,e^{i\theta}).
\end{equation}

As shown by Shraiman and Bensimon \cite{shraiman84}, Laplacian growth implies
that the conformal map has time dependence given by
\begin{equation}
\label{shr}
\frac{\partial g}{\partial t}=-i\frac{\partial g}{\partial \theta}L\left(\left|\frac{\partial g}{\partial
\theta}\right|^{-2}\right),
\end{equation}
where, if
\begin{equation}
\label{defdk}
\left|\frac{\partial g}{\partial \theta}\right|^{-2}=\sum_{k=-\infty}^{\infty}d_ke^{-ik\theta},
\end{equation}
then
\begin{equation}
L\left(\left|\frac{\partial g}{\partial \theta}\right|^{-2}\right)=d_0+2\sum_{k=1}^{\infty}d_ke^{-ik\theta}.
\end{equation}
Define new scaled variables
\begin{equation}
\label{scale1}
a_k=c_k/c_0,\quad\quad b_k=d_k/d_0
\end{equation}
and a new independent variable $s(t)$ by 
\begin{equation}
\label{scale2}
ds/dt=d_0.
\end{equation}
In terms of these variables, Eq. (\ref{shr}) becomes
\begin{equation}
\label{scaled}
\frac{da_k}{ds}=(k-2)a_k+2\sum_{j=0}^{k}(1-j)a_jb_{k-j}.
\end{equation}
Because of the rescaling in Eqs.(\ref{scale1}) and (\ref{scale2}), Eq. (\ref{scaled}) continues
to make sense even in the limit as singularities of the conformal map move onto the unit
circle.  For example, $|b_k|\leq1$ for all $k$, even though $d_k$ blows up.  We define the $b_k$'s,
in case there are singularities on the unit circle, to have their limiting values as the
singularities move onto the unit circle from the inside.  With this understanding, Eq. (\ref{scaled})
describes Laplacian growth, both singular and nonsingular, 
in terms of the scaled mapping function
\begin{equation}
\label{H}
H(s,z)=\frac{G[(t(s),z)]}{c_0[(t(s)]}=z\sum_{k=0}^\infty a_k(s)z^{-k}.
\end{equation}

In the singular case, in which all singularities of $H$ are on the unit circle, $H$ is a 
Schwarz-Christoffel map onto a degenerate polygon, and therefore its derivative has the form
\begin{equation}
\label{dHdz}
\frac{\partial H}{\partial z}=\prod_{j=1}^M(1-e^{i\beta_j}/z)^{\alpha_j} \prod_{k=1}^N(1-e^{i\gamma_k}/z)
\end{equation}
As a conformal map, $H$ has singularities at points on the unit circle which we have called $\beta_j$
and $\gamma_k$.  The image of an arc of the unit circle under $H$, so long as it does not
contain a singularity, is a straight line segment.  At the singularity $\beta_j$, however,
the image line turns through the angle $\alpha_j\pi$, which may be either positive (counterclockwise)
or negative (clockwise).  It is understood that $|\alpha_j|<1$.  
At the singularity $\gamma_k$ the image turns through the angle $\pi$,
i.e., the line retraces itself:  see Fig. 1.  The $\gamma$ singularity may seem to be only a special case
of the $\beta$ singularity, corresponding to $\alpha=1$,
but we have distinguished it because growth takes place entirely
at the $\gamma$ singularities:  the $\gamma$'s are different.
  (Notation:  $\alpha$=``angle''; $\beta$=``branch point''; $\gamma$=
``growth tip.''  The $\alpha_j$ of this paper was called $\alpha_j-1$ in \cite{peterson89}.)

The $\alpha_j$, $\beta_j$, and $\gamma_k$ are by no means arbitrary.  First, because the image
polygon turns through a total angle $2\pi$, by conformality, one must have, since there
are $M$ branchpoints and $N$ growth tips,
\begin{equation}
\label{sumangles}
N+\sum_{j=1}^{M}\alpha_j=2
\end{equation}
Second, because the image polygon looks like a tree graph, each edge of the image is traversed
twice, once in each direction.  This means that the integrals of $\partial H/\partial z$ along arcs
of the unit circle from one singularity to the next, which are singular integrals, must cancel
in pairs, an intricate condition on the positions of the $\beta$'s and $\gamma$'s.

Suppose at time $s=0$ we have such a conformal map $H$.
The equation of motion for $H$, Eq. (\ref{scaled}), should be recast as equations of 
motion for the singularities of $H$.  The $a_k$'s of Eq. (\ref{scaled}) are just the
coefficients in the power series for $H$, according to Eq. (\ref{H}), but we still need
the $b_k$'s. 
To compute the $b_k$'s, using Eqs. (\ref{defdk}) and (\ref{scale1}), we must Fourier
transform $|\partial H/\partial z|^{-2}$, restricted to the unit circle, with the
singularities displaced slightly inside.  The Fourier transform integrals are dominated
by the $\gamma$ singularities, and as the singularities move onto the unit circle, the
entire contribution comes from them.  Thus there is a simple formula for $b_k$,
\begin{equation}
b_k=\sum_{j=1}^N v_j e^{ik\gamma_j}.
\end{equation}
Here the ``weights'' $v_j$ are determined by
\begin{equation}
v_j \propto \lim_{z\rightarrow e^{i\gamma_j}}\left |\frac{\partial H/\partial z}{z-e^{i\gamma_j}}\right|^{-2}
\end{equation}
with the constant of proportionality determined by the normalization
\begin{equation}
\label{normalizev}
\sum_{j=1}^N v_j=b_0=1.
\end{equation}
Define the function
\begin{equation}
B(z)=\sum_{k=0}^\infty b_kz^{-k}=\sum_{j=1}^N\frac{v_k}{1-e^{i\gamma_j}/z}
\end{equation}
Then multiplying each side of Eq. (\ref{scaled}) by $z^{-k+1}$ and summing over $k$ gives
\begin{equation}
\label{dHds}
\frac{\partial H}{\partial s}=-H+(2B-1)z\frac{\partial H}{\partial z}
\end{equation}
It is $\partial H/\partial z$ rather than $H$ that we know explicitly, so
take the derivative of Eq. (\ref{dHds}) with respect to $z$.  The left side 
becomes
\begin{equation}
\frac{\partial^2 H}{\partial s \partial z}=-i\frac{\partial H}{\partial z}\left(
\sum_{j=1}^M\frac{\alpha_j e^{i\beta_j}/z}{1-e^{i\beta_j}/z}\,\frac{d\beta_j}{ds}+
\sum_{k=1}^N\frac{e^{i\gamma_k}/z}{1-e^{i\gamma_k}/z}\,\frac{d\gamma_k}{ds}\right )
\end{equation}
and the right side becomes an explicitly known expression.  One can now cancel
the factor $-i\partial H/\partial z$ in all terms.   Multiplying by $(1-e^{i\beta_j}/z)$ and
taking the limit as $z\rightarrow e^{i\beta_j}$ isolates $d\beta_j/ds$, and similarly
multiplying by $(1-e^{i\gamma_k}/z)$ and taking the limit as $z\rightarrow e^{i\gamma_k}$
isolates $d\gamma_k/ds$.  The result, after algebraic simplification, using Eqs. (\ref{normalizev})
and (\ref{sumangles}), is
\begin{eqnarray}
\label{dbetads}
\frac{d\beta_j}{ds}&=&\sum_{k=1}^{N}v_k \cot\left(\frac{\beta_j-\gamma_k}{2}\right) \\
\label{dgammads}
\frac{d\gamma_k}{ds}&=&v_k\sum_{j=1}^{M}\alpha_j \cot\left(\frac{\gamma_k-\beta_j}{2}\right)
	+\sum_{j\neq k}(v_k+v_j)\cot\left(\frac{\gamma_k - \gamma_j}{2}\right)
\end{eqnarray}
We can also note
\begin{equation}
\label{vk}
v_k=w_k/W
\end{equation}
where
\begin{eqnarray}
\label{wk}
w_k&=&\prod_{j=1}^M\sin^{-2\alpha_j}\left(\frac{\gamma_k-\beta_j}{2}\right )\prod_{j\neq k}\sin^{-2}
	\left(\frac{\gamma_k-\gamma_j}{2}\right) \\
\label{W}
W&=&\sum_{k=1}^{N}w_k
\end{eqnarray}
It is understood in Eq. (\ref{wk}) that $w_k$ is real and positive.  
Eqs. (\ref{dbetads})-(\ref{W})
represent the dynamics of singular Laplacian growth as an autonomous system of ODE's.

Remarkably, this system is a kind of gradient system:
\begin{eqnarray}
\frac{d\beta_j}{ds}&=&-\frac{1}{\alpha_j}\,\frac{\partial \ell n\, W}{\partial \beta_j}\\
\frac{d\gamma_k}{ds}&=&-\frac{\partial \ell n\,W}{\partial \gamma_k}
\end{eqnarray}
This is gradient flow in the space of parameters $(\beta_1,...,\beta_M,\gamma_1,...,\gamma_N)$
endowed with the metric tensor
\begin{equation}
\label{metric}
g=diag(\alpha_1,...,\alpha_M,1,...,1)
\end{equation}
Since, according to Eq. (\ref{sumangles}), the $\alpha$'s are negative, on average, if $N>2$,
this metric is indefinite.  The gradient flow is toward certain critical points of $\ell n\,W$ which
are not minima.  These critical points are the equilibria of singular Laplacian growth.  They
can be found by integrating the system of Eqs. (\ref{dbetads})-(\ref{W}).   Even if the starting
state does not satisfy all the conditions described after Eq. (\ref{sumangles}), it will still
approach a state which does satisfy them.  We describe these equilibrium states more precisely
below.  This stability of the flow, which is a familiar property of gradient flows, is an
indication that singular Laplacian growth is {\em not} sensitive to initial
conditions, contrary to what one might have expected,
and hence that the peculiar sensitivity of Laplacian growth in general has
either been lost in the passage to the singular case, or that it arises from some other cause.
We suggest below that that other cause is the non-uniqueness property of the system, 
still to be described.

The dynamics of singularities described by Eqs. (\ref{dbetads})-(\ref{W}) can be pictured
very simply.  The $\beta$'s are repelled by the $\gamma$'s on the unit circle, and the
$\gamma$'s repel each other.  The ``strength'' with which each $\gamma_k$ repels other singularities
is given by the corresponding $v_k$ (always $>0$).  The $\beta$'s, on the other hand, do not
interact directly with each other.  No singularity can pass through another one -- they
always keep the same order around the unit circle.  Those $\beta's$ between
any two adjacent $\gamma$'s are, however, driven by them toward some intermediate point
where, in effect, they coalesce into a single effective $\beta$, characterized by a 
single effective $\alpha$ which is the sum of all the contributing $\alpha's$. 
The way a $\beta$ singularity approaches its limit position is the way $x(s)$ approaches 0 in
\begin{equation}
  dx/ds=-x^2,
\end{equation}
namely
\begin{equation}
x(s)=x_0/(1+x_0s),
\end{equation}
that is, it takes infinite time.
By the second derivative test, there is only one equilibrium position
for the $\beta$'s between each pair of $\gamma$'s. Here all the $\beta$'s
will collect.  
Thus in the limit as $s\rightarrow\infty$
the equilibrium states of singular Laplacian growth have $\gamma$'s and $\beta$'s
alternating, and look like $N$ needles radiating from a single central point.  One
can even write a formula in closed form for $H$ in this case,
\begin{equation}
H=z\prod_{k=1}^{N}(1-e^{i\beta_k}/z)^{\alpha_k+1}
\end{equation}
where the $\beta$'s and $\alpha$'s are the effective ones.
We can also understand this outcome, in a more physical way, 
by realizing that the continual rescaling
means all internal structure shrinks to a point, leaving
only the growth tips as visible features.  What is not obvious from this
description, but is observed, is that typically some of the $\gamma$'s are
entrained with the $\beta$'s, and coalesce with them (where they contribute $+1$
to the effective $\alpha$).  This amounts to the scaling away of needles.
It turns out that 
the generic stable equilibria have $N\leq 3$.  If the initial configuration has
$N>3$, and is the least bit asymmetrical,
some of the growing tips lose out in the competition to grow,
and disappear as $s\rightarrow\infty$, leaving only three
(or fewer) needles in the limit.
 
This result might appear puzzling, since it seems to imply that Laplacian
growth should be a process of {\em simplification}, contrary to
the increasing complexity which is
observed, and which is the whole motivation for studying it.  
That puzzle is resolved in the next section.

\section{Non-uniqueness of the Dynamics}
If one reverses the sense of time and integrates the system backwards,
the repelling character of the $\gamma$'s becomes an attraction.
In particular,two $\beta$'s adjacent on either side of a $\gamma$ may be attracted to it,
move toward it,
and coalesce with it, essentially annihilating, leaving just the $\gamma$.
  Unlike the coalescence described at the end of
the previous section, which takes infinite time,
this coalescence occurs {\em in finite time}.
Roughly, one can estimate from Eq. (\ref{dbetads}) that a $\beta$ approaches
a nearby $\gamma$ the way x(s) approaches 0 in
\begin{equation}
dx/ds=1/x,
\end{equation}
(integrating backward from $s=0$),
namely 
\begin{equation}
x(s)=\sqrt{x_0^2+2s}.
\end{equation}
At times earlier than the coalescence time $-x_0^2/2$,
the $\beta$ singularities were simply not present.
If one now examines this solution to the system with the usual forward
sense of time, one sees, at some arbitrary time -$x_0^2/2$, two $\beta$ singularities
suddenly produced on either side of a $\gamma$, which hadn't
been there before.  To satisfy Eq. (\ref{sumangles}), the $\alpha$'s which
characterize these $\beta$'s must add to zero.  The geometrical effect
of this process is that a kink of deviation angle $\alpha$ suddenly
appears in the growing needle represented
by $\gamma$, like the kink shown in Fig. 1, which might have
formed from a single straight needle.  This kinking may happen at any arbitrary time.
A more careful argument (in the Appendix) says that
if a kink forms at $\gamma$ at $s=0$, the leading behavior in the motion
of singularities is
\begin{eqnarray}
\label{sing1}
(\gamma-\beta_1)&\propto&\sqrt{\frac{1+\alpha_1}{1+\alpha_2}}\,s^{1/2} \\
(\gamma-\beta_2)&\propto&-\sqrt{\frac{1+\alpha_2}{1+\alpha_1}}\,s^{1/2} 
\end{eqnarray}
with $\alpha_1=-\alpha_2$.

In addition, a second kind of coalescence is seen in backward integration, in which
two $\beta$'s, with angles $\alpha_1$ and $\alpha_2$ 
on either side of a $\gamma$, coalesce to leave a single
$\beta$ with angle $1+\alpha_1+\alpha_2$.  
This happens only if $1+\alpha_1+\alpha_2>0$ and $\alpha_1+\alpha_2<0$.
Geometrically it corresponds to the shrinking away of a
needle in finite time (the growth tip $\gamma$ is lost), on the outside
of a corner of angle $1+\alpha_1+\alpha_2$.  What it means in forward integration is that at
any time a needle may begin {\em growing} on the outside of a corner, as
in the process which takes Fig. 1 to Fig. 2.
The motion of singularities in this case, in leading order, is
\begin{eqnarray}
\gamma-\beta_1\propto\sqrt{\frac{1+\alpha_1}{1+\alpha_2}}\,s^{1/2(1+\alpha_1+\alpha_2)} \\
\label{sing4}
\gamma-\beta_2\propto-\sqrt{\frac{1+\alpha_2}{1+\alpha_1}}\,s^{1/2(1+\alpha_1+\alpha_2)} 
\end{eqnarray}
with $1+\alpha_1+\alpha_2>0$, $\alpha_1+\alpha_2<0$.
(In Ref. \cite{peterson89} the factor $1+\alpha_1+\alpha_2$
in the exponent's denominator mistakenly appeared
in the numerator.)

\section{Discussion}
The observations of Section III mean that the system of ODE's Eqs. (\ref{dbetads})-(\ref{W}),
although it looks unremarkable, has the peculiar property that its solutions are
highly non-unique.  New singularities can appear by the above
two elementary processes at any time.  
In combination one has more complicated processes:
a kink followed by a new needle at the outside of the new corner amounts to
tip splitting, for example, and this can happen at any time.  The
equilibria described in Section II are never attained if such 
processes, which are allowed by the differential equation, continually intervene.  
Thus singular Laplacian growth supports complex non-equilibrium
behavior after all.
  
It is interesting to see what the model looks like if one integrates it forward,
introduces new singularities, integrates again, adds more singularities, etc.
Examples are shown in Figs. (3) and (4), where symmetrical tip splitting was introduced at 
intervals of 0.1 time unit.
To interpret the evolving positions of singularities in terms of 
the corresponding image region $D$,
which is what is shown, it was necessary to integrate Eq. (\ref{dHdz}) numerically.
Each edge is represented by a singular integral.  These integrals were done by 
Gauss-Jacobi integration, as described by
L.N. Trefethen in Ref. \cite{trefethen}.  The accumulating error in
these numerical integrals, as one steps along each edge to the next, especially in
light of the usual sensitivity of numerical conformal maps, might have been
expected to produce nonsensical pictures, but in fact the numerical error (failure
to retrace edges accurately) is
just barely visible in these examples.  
(Eventually, of course, the accumulating error does become
large, but the good numerical behavior of the system again makes the point
that singular Laplacian growth is {\em not} particularly sensitive to error or noise.
Its sensitivity to perturbations comes entirely through the non-uniqueness property.)

\section{Relation to Other Work}
Most of those who have used the conformal mapping method have followed
Shraiman and Bensimon \cite{shraiman84}
in restricting the derivative of the conformal map $H$ to
be a rational function.
From some points of view this is a rather drastic restriction
on the analytic structure of $H$.
Whether it is a good enough representation of $H$ to learn the
full implications of the conformal mapping method is not clear.
Arguments that the boundary value of $H'$ can
always be approximated by the boundary value of a rational function are
not very convincing in a context where it is precisely the nature of the singularities
which is the basis of the theory.  It had already been noticed in Ref. \cite{peterson89} that
branch points play an essential role in the singular theory.  Nonetheless, an interesting 
comparison between the singular case and the rational case
is possible.

An example is Ref. \cite{blumenfeld95}, in which R. Blumenfeld and R.C. Ball invent
a mechanism of ``particle creation'' (i.e singularity creation)
to model tip splitting.  
In their model, since $H'$ is rational, the only singularities are the zeros and
poles of $H'$.  The mechanism they propose is that a zero creates a second zero and
a pole.  The two zeros represent the two growing tips after the split, and the pole
represents the division between them.  

Tip splitting in the singular theory, as described in Ref. \cite{peterson89}
and in this paper, does not have to be invented --
it is naturally and unavoidably part of the theory -- and it looks slightly more complicated:  
a $\gamma$ gives rise to three
$\beta$'s and another $\gamma$ (as in Figs. 1 and 2).  
But in fact this amounts to the same thing.  The 
resulting two $\gamma$'s
are zeros of $H'$, and, by the geometry of the situation,
the three new $\alpha$'s add to $-1$.  This means that the
three $\beta$'s, from a distance much greater than their mutual separation, 
look like a pole of $H'$ (see Eq. (\ref{dHdz})).  
The mechanism proposed by Blumenfeld and Ball is thus a
kind of smeared version of the singular mechanism, already described
in Ref. \cite{peterson89}

It is especially remarkable that Blumenfeld and Ball invented their mechanism
entirely on the basis of physical phenomenology, and were unaware of Ref. \cite{peterson89}.
Their mechanism of particle creation, although it is {\em ad hoc},
corresponds as precisely as it could have to the {\em only}
mechanism in the singular theory for non-trivial dynamics.  This suggests
that the singular theory is close enough to real phenomenology to be useful,
and that it does retain the essential features of Laplacian growth.

\section{Generalities}
To focus on the details of singular Laplacian growth is, to some extent, to sidestep
a much bigger question:  what is going on here with non-uniqueness?  
Aren't differential equation supposed
to have unique solutions?  We all know textbook examples where uniqueness fails,
but the failure occurs on some small set, and for equations which wouldn't
arise in physics.  Here are equations which arise in a system which has been much
discussed in physics, and uniqueness fails for every solution at every time.
The least one can say is that the equations of motion do not determine the motion,
even in the small.

I believe this is actually mathematical {\em terra incognita}.  Such equations
do not even have a name.  How would one characterize them generally?  Are they
in some sense common, or are they rare?  I think of calling them ``fragile
differential equations,'' because, at least in this example, the non-uniqueness
arises by the tendency of singularities to ``break apart,'' but perhaps a
more general understanding would reveal that this name is somehow misleading.
``Fragile'' sounds a little bit like ``fractal,'' but is not the same, another
reason I like the name.

On a more physical level, what does it mean for a physical system if it is
described by equations which, in some limit, become ``fragile''?  A fragile
system does not fully determine the evolution, but it does restrict it.
What is the nature of the restriction?  
These seem like good questions for
the future.

\section*{Appendix}
We derive Eqs. (\ref{sing1})-(\ref{sing4}), the leading
behavior of singularities $\beta_1$, $\gamma$, $\beta_2$ when they
are very close to each other (in that order), and not close
to other singularities.  Let $\alpha_1$ and $\alpha_2$ be
the corresponding angle parameters, and $v$ the ``strength'' of $\gamma$.  
According to Eqs. (\ref{dbetads})-(\ref{dgammads}),
keeping only the most singular terms, in leading order they obey
\begin{eqnarray}
\frac{d\beta_1}{ds}&=&\frac{2v}{\beta_1-\gamma} \\
\frac{d\beta_2}{ds}&=&\frac{2v}{\beta_2-\gamma}\\
\frac{d\gamma}{ds}&=&\frac{2v\alpha_1}{\gamma-\beta_1}+\frac{2v\alpha_2}{\gamma-\beta_2}
\end{eqnarray}
Let
\begin{eqnarray}
P&=&\gamma-\beta_1\\
Q&=&\gamma-\beta_2.
\end{eqnarray}
Then, subtracting,
\begin{eqnarray}
\label{dPds}
\frac{dP}{ds}&=&\frac{2v(1+\alpha_1)}{P}+\frac{2v\alpha_2}{Q}\\
\label{dQds}
\frac{dQ}{ds}&=&\frac{2v\alpha_1}{P}+\frac{2v(1+\alpha_2}{Q}
\end{eqnarray}
Dividing, we have the homogeneous equation
\begin{equation}
\frac{dP}{dQ}=\frac{(1+\alpha_1)Q+\alpha_2P}{\alpha_1Q+(1+\alpha_2)P}
\end{equation}
which separates when written in terms of the variable $P/Q$.  The complete solution,
in implicit form,
is
\begin{equation}
(P-Q)^{\nu_1}
	[(1+\alpha_2)P+(1+\alpha_1)Q]^{\nu_2}=const.
\end{equation}
where
\begin{eqnarray}
\nu_1&=&(1+\alpha_1+\alpha_2)/(2+\alpha_1+\alpha_2)\\
\nu_2&=&1/(2+\alpha_1+\alpha_2)
\end{eqnarray}
Since this is supposed to hold as $P$, $Q$ approach zero, the only relevant
value of the constant is zero.  The solution $P=Q$ is not relevant to
this situation, since $P$ and $Q$ must have opposite sign.  Thus
\begin{equation}
\label{PQ}
(1+\alpha_2)P+(1+\alpha_1)Q=0.
\end{equation}
Using Eqs. (\ref{vk})-(\ref{W}) together with the fact, found in Eq. (\ref{PQ}),
that $\gamma-\beta_1$ and $\gamma-\beta_2$ are simply proportional, we see that
$v$ is non-singular if $\alpha_1+\alpha_2\geq 0$, and
\begin{equation}
v \sim P^{-2\alpha_1-2\alpha_2}
\end{equation}
if $\alpha_1+\alpha_2<0$.  Thus from Eq. (\ref{dPds}), 
\begin{equation}
dP/ds\sim P^{-1}
\end{equation} 
if
$\alpha_1+\alpha_2\geq 0$, as in the rough argument of Section III, and
\begin{equation}
\frac{dP}{ds}\sim P^{-2\alpha_1-2\alpha_2-1}
\end{equation}
if $\alpha_1+\alpha_2<0$.  These results are summarized in Eqs. (\ref{sing1})-(\ref{sing4}).
\end{multicols}


\newpage
\begin{figure}[htb]
\centerline
{
\epsfxsize=12cm
\epsfbox{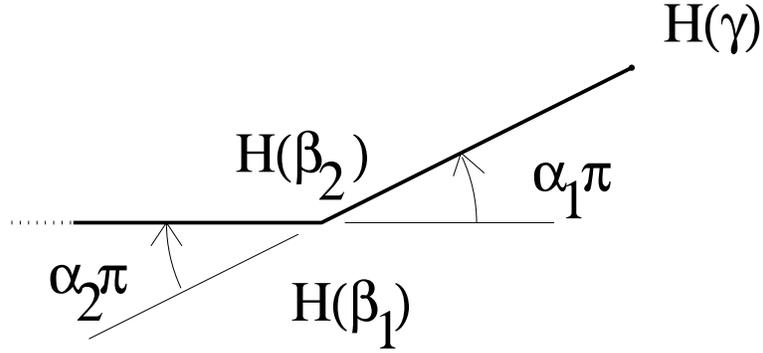}
}
\caption{
The image under a degenerate Schwarz-Christoffel map of an arc containing
singularities ...$\beta_1$, $\gamma$, $\beta_2$,...  The image comes in
from the left, turns through an angle $\alpha_1\pi$ at $H(\beta_1)$, stops
and reverses direction at $H(\gamma)$, turns through an angle $\alpha_2\pi$ at
$H(\beta_2)$, and exits on the left.  In this example, showing a kink in
a growing needle, $\alpha_1=-\alpha_2$.
}
\end{figure}
\begin{figure}[htb]
\centerline
{
\epsfxsize=12cm
\epsfbox{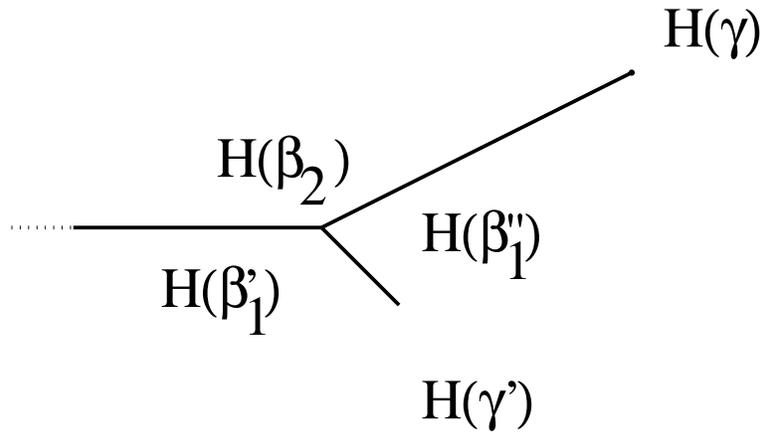}
}
\caption
{
A new needle may grow on the outside of the corner in Fig. 1.
Here the singularity $\beta_1$ has split into two branch points,
$\beta_1'$ and $\beta_1$'', and a new growth tip $\gamma$'.
}
\end{figure}
\begin{figure}[htb]
\centerline
{
\epsfxsize=12cm
\epsfbox{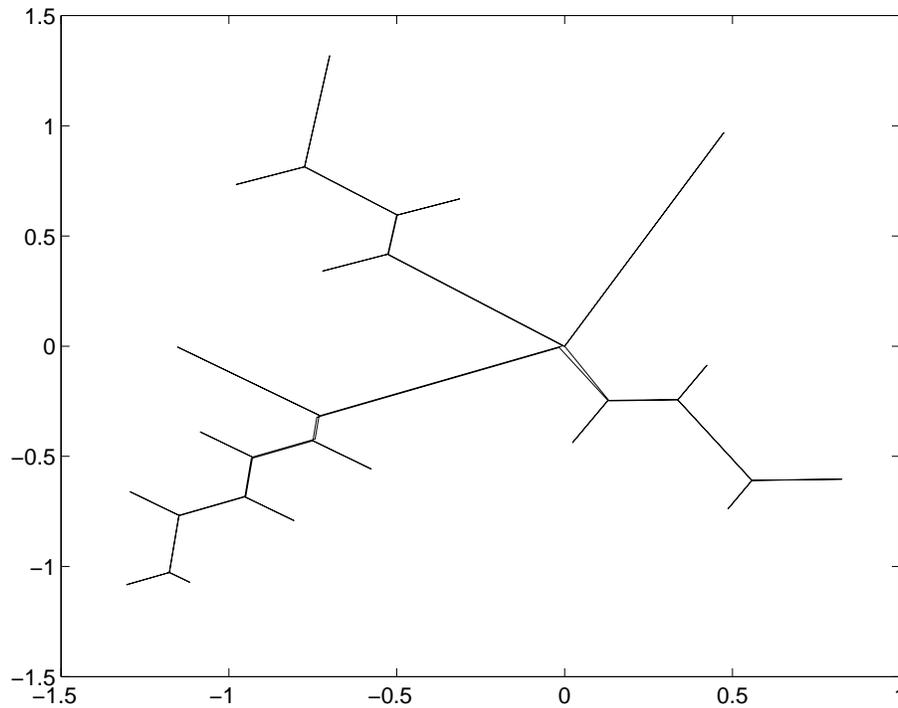}
}
\caption{
At each time interval $\Delta s=0.1$ the growing tip with largest strength is split.
The initial configuration was four random needles radiating from a point, but the
growth law is completely deterministic.
}
\end{figure}
\begin{figure}[htb]
\centerline
{
\epsfxsize=12cm
\epsfbox{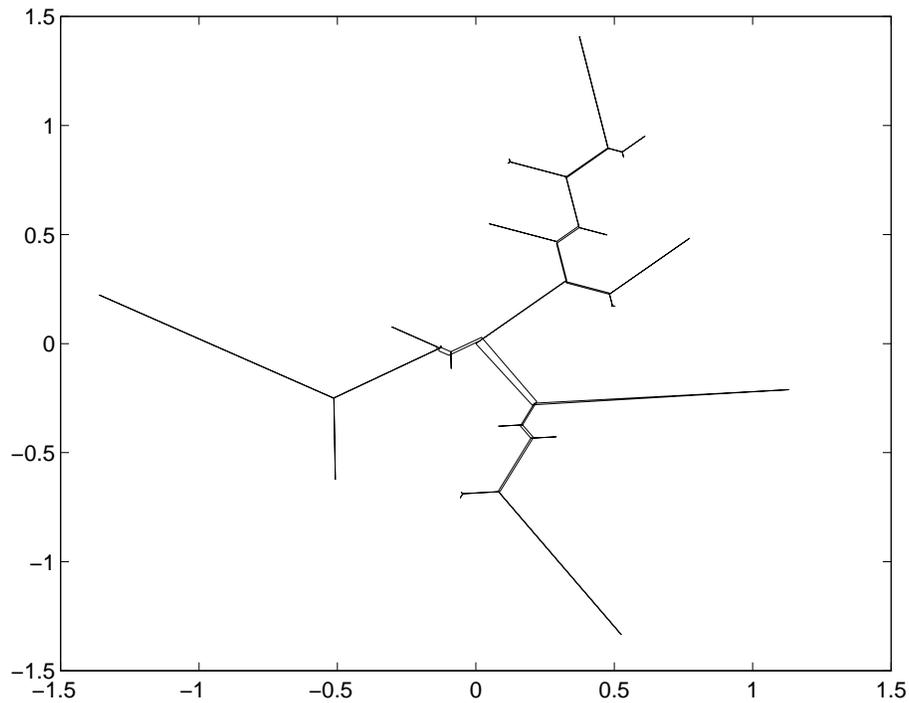}
}
\caption{
At each time interval $\Delta s=0.1$ a growing tip is randomly selected and split.
The probability of a tip's being selected is proportional to its strength $v$.
The initial configuration was three random needles radiating from a point. 
}
\end{figure}
\end{document}